\documentstyle[12pt,epsfig  ]{article}
\begin{document}

\def\ie{\hbox{\it i.e.}{}}      \def\etc{\hbox{\it etc.}{}}
\def\eg{\hbox{\it e.g.}{}}      \def\cf{\hbox{\it cf.}{}}
\def\etal{\hbox{\it et al.}}    \def\vs{\hbox{\it vs.}{}}
\def\dash{\hbox{---}}
\newcommand\refq[1]{$^{#1}$}
\def\fb{~{\rm fb}}
\def\pb{~{\rm pb}}
\def\ev{\,{\rm eV}}
\def\mev{\,{\rm MeV}}
\def\gev{\,{\rm GeV}}
\def\tev{\,{\rm TeV}}
\def\wh{\widehat}
\def\wt{\widetilde}
\def\abs#1{\left| #1\right|}   

\relax
\newcommand{\slsh}{\rlap{$\;\!\!\not$}}     
\newcommand{\as}{{\ifmmode \alpha_S \else $\alpha_S$ \fi}}
\newcommand{\ep}{\epsilon}
\newcommand{\be}{\begin{equation}}
\newcommand{\ee}{\end{equation}}
\newcommand{\bea}{\begin{eqnarray}}
\newcommand{\eea}{\end{eqnarray}}
\newcommand{\ba}{\begin{array}}
\newcommand{\ea}{\end{array}}
\newcommand{\bi}{\begin{itemize}}
\newcommand{\ei}{\end{itemize}}
\newcommand{\bn}{\begin{enumerate}}
\newcommand{\en}{\end{enumerate}}
\newcommand{\bc}{\begin{center}}
\newcommand{\ec}{\end{center}}
\newcommand{\ul}{\underline}
\newcommand{\ol}{\overline}
\newcommand{\sm}{${\cal {SM}}$}
\newcommand{\mssm}{${\cal {MSSM}}$}
\newcommand{\Dir}{\kern -6.4pt\Big{/}}
\newcommand{\Dirin}{\kern -10.4pt\Big{/}\kern 4.4pt}
\newcommand{\DDir}{\kern -7.6pt\Big{/}}
\newcommand{\DGir}{\kern -6.0pt\Big{/}}

\def\Ord{\buildrel{\scriptscriptstyle <}\over{\scriptscriptstyle\sim}}
\def\OOrd{\buildrel{\scriptscriptstyle >}\over{\scriptscriptstyle\sim}}

\def\ap#1#2#3{
        {\it Ann. Phys. (NY) }{\bf #1} (19#3) #2}
\def\app#1#2#3{
        {\it Acta Phys. Pol. }{\bf #1} (19#3) #2}
\def\cmp#1#2#3{
        {\it Commun. Math. Phys. }{\bf #1} (19#3) #2}
\def\cpc#1#2#3{
        {\it Comput. Phys. Commun. }{\bf #1} (19#3) #2}
\def\ijmp#1#2#3{
        {\it Int .J. Mod. Phys. }{\bf #1} (19#3) #2}
\def\ibid#1#2#3{
        {\it ibid }{\bf #1} (19#3) #2}
\def\jmp#1#2#3{
        {\it J. Math. Phys. }{\bf #1} (19#3) #2}
\def\jetp#1#2#3{
        {\it JETP Sov. Phys. }{\bf #1} (19#3) #2}
\def\ib#1#2#3{
        {\it ibid. }{\bf #1} (19#3) #2}
\def\mpl#1#2#3{
        {\it Mod. Phys. Lett. }{\bf #1} (19#3) #2}
\def\nat#1#2#3{
        {\it Nature (London) }{\bf #1} (19#3) #2}
\def\np#1#2#3{
        {\it Nucl. Phys. }{\bf #1} (19#3) #2}
\def\npsup#1#2#3{
        {\it Nucl. Phys. Proc. Sup. }{\bf #1} (19#3) #2}
\def\pl#1#2#3{
        {\it Phys. Lett. }{\bf #1} (19#3) #2}
\def\pr#1#2#3{
        {\it Phys. Rev. }{\bf #1} (19#3) #2}
\def\prep#1#2#3{
        {\it Phys. Rep. }{\bf #1} (19#3) #2}
\def\prl#1#2#3{
        {\it Phys. Rev. Lett. }{\bf #1} (19#3) #2}
\def\physica#1#2#3{
        { Physica }{\bf #1} (19#3) #2}
\def\rmp#1#2#3{
        {\it Rev. Mod. Phys. }{\bf #1} (19#3) #2}
\def\sj#1#2#3{
        {\it Sov. J. Nucl. Phys. }{\bf #1} (19#3) #2}
\def\zp#1#2#3{
        {\it Z. Phys. }{\bf #1} (19#3) #2}
\def\tmf#1#2#3{
        {\it Theor. Math. Phys. }{\bf #1} (19#3) #2}

\begin{flushright}
{\large ETH--TH/97--11}\\
{\rm March 1997\hspace*{.5 truecm}}\\
\end{flushright}

\vspace*{2truecm}

\begin{center}
{\Large \bf 
Theoretical  Aspects of Higgs Hunting at LHC
\footnote{Talk given at the Ringberg Workshop on ``The Higgs puzzle -
  What can we learn from LEP2, LHC, NLC, and FMC?''
 (8 -- 13 December,1996)\,. \\[0.5cm] 
E-mails:
 kunszt@itp.phys.ethz.ch} }\\[0.5cm]
{\large 
Zoltan Kunszt}\\[0.15 cm]
{\it  Institute of Theoretical Physics, ETH, Z\"urich, Switzerland}
\end{center}
\vspace*{5truecm}
\begin{abstract}
{\noindent 
I review recent developments  in the theoretical 
study  of Higgs search at LHC.
}
\end{abstract}
\newpage
 \normalsize\baselineskip=15pt
\setcounter{footnote}{0}
\renewcommand{\thefootnote}{\alph{footnote}}
\section{Introduction}
One of the most important physics goal of the LHC is  the search for
 Higgs boson(s)~\cite{guide,LHC}.
During the last six years the design energy and machine luminosity
 have been changed (not always so as we hoped).
 For the simulation works carried out during
 the Aachen workshop (1990) $\sqrt{s}$
 $=$ $16\tev$
energy and $\Delta{\cal L}$ $=$ $10^5\pb^{-1}$ - $3\times
10^5\pb^{-1}$ integrated luminosity per year
was assumed.    
The CMS and ATLAS technical proposals~\cite{ATLAS,CMS} (1994) used $\sqrt{s}$
 $=$ $14\tev$ and $\Delta{\cal L}$ $\le$ $10^5\pb^{-1}$. These values were
 approved by the CERN Council in 1994 December but it could
not be excluded that in the  first year of operation 
the experiments start data taking at  $\sqrt{s}$
 $=$ $10\tev$ and $\Delta{\cal L}$ $\le$ $10^4\pb^{-1}$.
As a result of these changes most
 of the numerical results to be found in the literature
are now out-of-date. This motivates 
the   update of the old calculations~\cite{oldpaper}.
New theoretical and experimental information also calls for a
re-examination of the bench-mark values of the Standard Model Higgs
Higgs branching ratios and production rates at the LHC~\cite{MWJSZK,JGSWS}.

According to the so called 
``{\it no loose scenario}'' in searching for the mechanism of the
electroweak symmetry breaking 
\begin{itemize}
\item[i)]
either the Higgs will be discovered at LEP2 or LEP2 will establish
a lower limit of $m_H>95-100\gev$;
\item[ii)]
 it is expected that LEP1 and SLC
will confirm an upper limit $600\gev>m_H$ by the end of the present
LEP2, SLC, LEP2 and Tevatron program;
\item[iii)] 
LHC will be able to discover the Standard Model Higgs boson in the
remaining interval $600\gev > m_H > 95\gev $.
\end{itemize}
Other options are also possible. For example 
 if the Higgs boson will be discovered at LEP2 
LHC can put stringent test on the supersymmetrised Standard Model;
or if  at LHC  the Standard Model Higgs with mass $m_H> 160\gev$ will
be found the Minimal Supersymmetric Standard Model will be ruled out;
or  LHC may
find evidence for supersymmetry etc.
This is exciting program require an unprecedented experimental effort and
it has to be supported by a continuous improvement of the theoretical
signal and background calculations.
In a  recent systematic study
D. Froidevaux \etal~\cite{richter}
carried out   an extended simulation
work for the ATLAS detector. They argued that
since we do not know all QCD radiative correction
factors we should ignore all of them and so they used for
all signal and background processes the lowest order cross-section
formulas.  Their study covers all the  previously studied cross section
and background processes~\cite{oldpaper}.
While this is  a consistent approach 
one should not abandon the careful evaluation of the
very important higher order QCD corrections.
 M. Spira and collaborators prepared two very useful
 program packages~\cite{spirahgludec}
which include all the existing NLO corrections.
HDECAY  generates all branching fractions of the
Standard Model Higgs boson and the Higgs bosons of the 
Minimal Supersymmetric Standard Model (MSSM) while HGLUE
 provides the production 
cross sections of  the SM and MSSM Higgs bosons via gluon fusion
including  the NLO QCD corrections. In a new 
 interesting work  Kr\"amer \etal\ 
attempted to improve the description of the soft gluon 
radiation effects and also tried to estimate the size of the NNLO
hard scattering cross sections~\cite{spirasoft}.
Many additional improvements appeared recently to further analyze
the signals of the Higgs bosons of the MSSM~\cite{mssmall}\,.

\section{Electroweak and QCD input parameters}
It is of interest
to improve the 
 calculations of the  relevant cross sections and branching ratios
including recently calculated QCD next-to-leading order corrections,
new parton distributions fitted to recent HERA structure function data,
and new values for electroweak input parameters, in particular 
for the top quark mass. In view of future plans
cross sections  have to be calculated at two collider
energies, $\sqrt{s} = 10$~TeV and $14$~TeV.
 Detailed studies (see, for example, Refs.~\cite{guide,LHC})
 have shown that there is no single
production mechanism or decay channel which dominates the phenomenology
over the whole of the relevant Higgs mass range, $O(100\ {\mathrm{GeV}}) <
M_H < O(1\ {\mathrm{TeV}})$, rather there are several different scenarios
depending on the value of $M_H$.
The recent updates~\cite{MWJSZK,spirahgludec,JGSWS}  took into account that
\begin{itemize}
\item[{(i)}] next-to-leading order corrections are now known for most
of the subprocess production cross sections and partial decay widths;
\item[{(ii)}]  knowledge of the parton distribution functions has
improved as more precision deep inelastic and other data have become
available;
\item[{(iii)}] the range of possible input parameter values (in particular the
top quark mass $m_t$) has decreased as a result of precision
measurements from LEP, the Tevatron $p \bar p$ collider and other
experiments.
\end{itemize}
\noindent
 New  benchmark results are available for cross
sections and event rates as a function of $M_H$, for the two
`standard' LHC collision energies, $\sqrt{s} = 10$ and $14\ {\mathrm{TeV}}$.
The    Higgs production
cross sections, event rates and significance factors
used in Refs.~\cite{ATLAS} and \cite{CMS} have to be renormalized to the
most up-to-date values.
(Detailed of 
signals, backgrounds and search strategies
 can be found in the recent ATLAS~\cite{ATLAS} and CMS~\cite{CMS}
Technical Proposals (see also ref.~\cite{JGSWS}).
The latest  input parameters  
 are as follows
$$M_Z=91.186~{\mathrm {GeV}},\quad\quad \Gamma_Z=2.495~{\mathrm {GeV}},$$  
$$M_W=80.356~{\mathrm {GeV}},\quad\quad \Gamma_W=2.088~{\mathrm {GeV}},$$ 
\be\label{ewparam}
G_F=1.16639\times10^{-5}~{\mathrm {GeV}}^{-2},
\quad\quad\alpha_{em}\equiv \alpha_{em}(M_Z)= 1/128.9 .
\ee
Each of these parameters has a small measurement error, 
the effect of these on the event rates  is negligible
compared to other uncertainties.
The charged and neutral weak fermion--boson couplings are defined by
\be\label{GF}
g_W^2 = \frac{e^2}{\sin^2\theta_W} =  4 \sqrt{2} G_F M_W^{2}, \qquad
g_Z^2 = \frac{e^2}{\sin^2\theta_W\; \cos^2\theta_W} =  4 \sqrt{2} G_F M_Z^{2}.
\ee
For the vector and axial couplings
of the $Z$ boson to fermions, we should use the `effective leptonic' value
\be\label{s2w_eff}
\sin^2_{\rm {eff}}(\theta_W) = 0.232\,.
\ee
The QCD strong coupling enters explicitly in the
 production cross sections and in the branching ratios,
 and implicitly in the parton distributions. 
Since most quantities  are known to next-to-leading order,
it is more consistent to use two loop    $\alpha_s$, with
 $\Lambda^{(4)}_{\overline{{MS}}}=230$~MeV  
to match  the MRS(A) default parton distribution sets,
The scale for production cross sections is best  given by the Higgs mass
 while for the branching 
ratios one should use the  the prescriptions of
Refs.~\cite{SMHVp,SMHggQCDcorr}.
I recall that we need also the values of  the fermion masses
$m_\mu=0.105$~GeV, $m_\tau=1.78$~GeV, $m_s=0.3$~GeV, $m_c = 1.4$~GeV,
 $m_b=4.25$~GeV
and $m_t=175$~GeV .
The first generation of fermions and all neutrinos are taken to be
massless, i.e. $m_u=m_d=m_e=m_{\nu_e}=0$ and
$m_{\nu_\mu}=m_{\nu_\tau}=0$, with zero decay widths.

It is of interest to study of the variation of the production
cross sections with $m_t$ in the range $165 < m_t < 185$~GeV, which subsumes
the recent direct measurement (CDF and D0 combined)
from the Tevatron $p \bar p$ collider
of $m_t = 175 \pm 6$~GeV.

Assuming   that $M_H =  90$~GeV is a 
 discovery mass limit for LEP2, one should cover 
 the mass range $90\ {\mathrm{GeV}} \leq M_H \leq 1\ {\mathrm{TeV}}$.
The discussion naturally falls into classes, depending on whether
$M_H$ is less than or greater than $O(2 M_W)$\footnote{This defines
the so-called `intermediate-mass' and `heavy' Higgs mass ranges,
respectively.}.
\newpage
\section{\sm\ Higgs branching ratios}
\label{sec:br}

 \begin{figure}[hbtp]
\vspace{8.cm}
\begin{picture}(7,7)(0, -1000)
\includegraphics{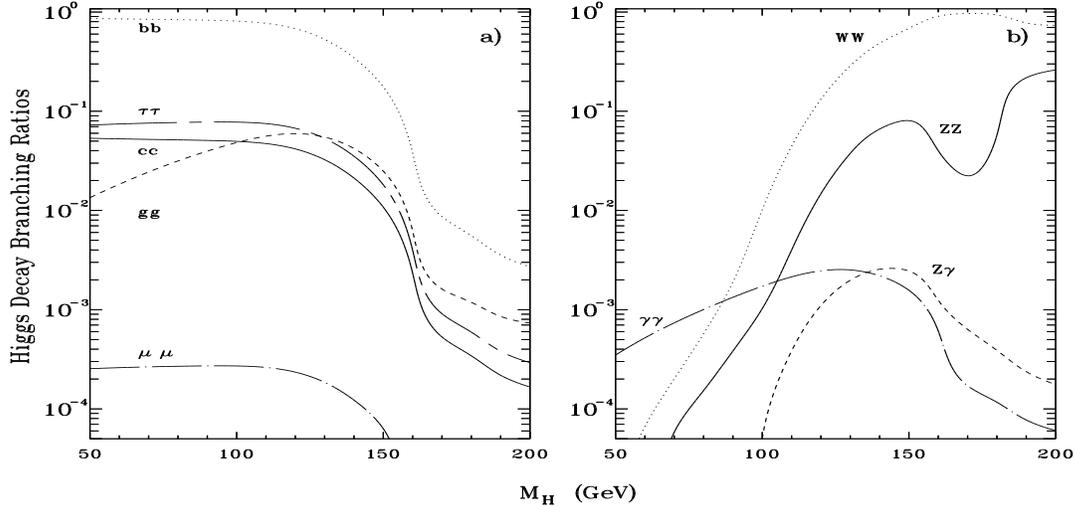}
\end{picture}
\caption{Branching ratios of the \sm\ Higgs boson in the mass
range 50~GeV $< M_H <$~200 GeV, for the decay modes: a) $b\bar b$,
$c\bar c$, $\tau^+\tau^-$, $\mu^+\mu^-$ and $gg$; b) $WW$,
$ZZ$, $\gamma\gamma$ and $Z\gamma$}
\label{fig:brlo}
\end{figure}
\noindent
The branching ratios of the \sm\ Higgs boson have been 
studied in many papers. 
A useful compilation of the early works on this subject can be found in
Ref.~\cite{guide}, where all the most relevant formulae
for on-shell decays are summarized.
Higher-order corrections
to  most  of the decay processes have also been computed
(for
up-to-date reviews  see Refs.~\cite{corrreview} 
and references therein), 
as well as 
the rates for the off-shell decays $H\to W^{*}W^{*},Z^{*}Z^{*}$,
 $H\to Z^{*}\gamma$ and $H\to t^*\bar t^*$.
 Threshold effects due to the possible
formation of $t\bar t$ bound states in the one-loop induced process
$H\to \gamma\gamma$ have also been studied \cite{threshold}. 

The updated calculations  include   the large 
QCD corrections to the  \sm\ Higgs partial widths into heavy quark 
pairs  and
into $Z\gamma$, $\gamma\gamma$ and $gg$ 
\cite{SMHggQCDcorr}.

The bulk of the QCD corrections to $H\to q\bar q$ can 
be absorbed into a `running' quark mass $m_q(\mu)$, evaluated at the energy 
scale $\mu=M_H$ (for example). 
The importance of this effect for the case $q=b$, 
with respect to intermediate-mass Higgs searches at the LHC, 
has been discussed already in Ref.~\cite{oldpaper}.
For sake of illustration, results on the Higgs branching ratios are summarized in 
Figs.~1 which
 shows the branching ratios, for $M_H \leq 200$~GeV,
for the channels: (a) $b\bar b$,
$c\bar c$, $\tau^+\tau^-$, $\mu^+\mu^-$ and $gg$; and (b) $WW$,
$ZZ$, $\gamma\gamma$ and $Z\gamma$. 
The patterns of the various curves are not significantly different
from those presented in Ref.~\cite{oldpaper}. The
inclusion of the QCD corrections in the quark-loop induced
decays (which apart from small changes in the parameter values
is the only
significant difference with respect to the calculation in \cite{oldpaper})
turns out to give a variation of at most a few per cent for the
decays $H\to\gamma\gamma$
and $H\to Z\gamma$, while for $H\to gg$ differences are of order
50--60\%. However this has little phenomenological relevance,
since this decay width makes a negligible contribution to the total width,
and is an unobservable channel in practice.
Note that for the below-threshold
decays $H\to W^*W^*$ and $H\to Z^*Z^*$, 
one integrates numerically over 
the virtualities of both
decay products, see for example Ref.~\cite{MSSMpaper},
thus avoiding errors in the threshold region.

\begin{figure}[hbtp]
\vspace{8.5cm}
\begin{picture}(7,7)
\includegraphics{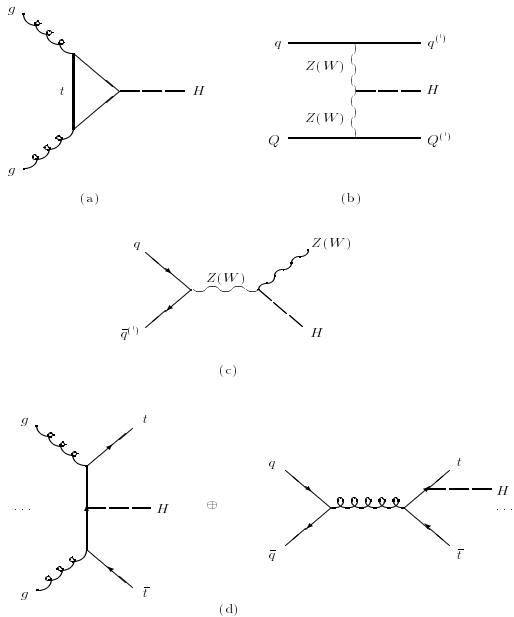}
\end{picture}
\caption{Feynman diagram for various production mechanisms}
\label{fig:diag}
\end{figure}

\section{\sm\ Higgs production cross sections and event rates}
\label{sec:cross}
 
There are only a few Higgs production mechanisms which lead to detectable 
cross sections at the LHC. Each of them makes use of the preference of
the \sm\ Higgs to couple to heavy particles: either massive vector
bosons ($W$ and $Z$) or massive quarks (especially $t$-quarks).
They are (see Fig.~2):
\begin{itemize}
\item[{(a)}]  gluon-gluon fusion \cite{xggh}, 
\item[{(b)}]  $WW$, $ZZ$ fusion \cite{xvvh},
\item[{(c)}]  associated production with $W$, $Z$ bosons \cite{xvh},
\item[{(d)}]  associated production with $t\bar t$ pairs \cite{xtth}.
\end{itemize}
A complete review on the early literature on $pp$ collider
 \sm\ Higgs boson phenomenology, based on these production mechanisms, 
 can be found in  Ref.~\cite{guide}.
Again for illustrative purpose we show the total cross-section values
for $\sqrt{s}=10\tev$ in Fig.3. 
 \begin{figure}[hbtp]
\vspace{7.5cm}
\begin{picture}(7,7)
\includegraphics{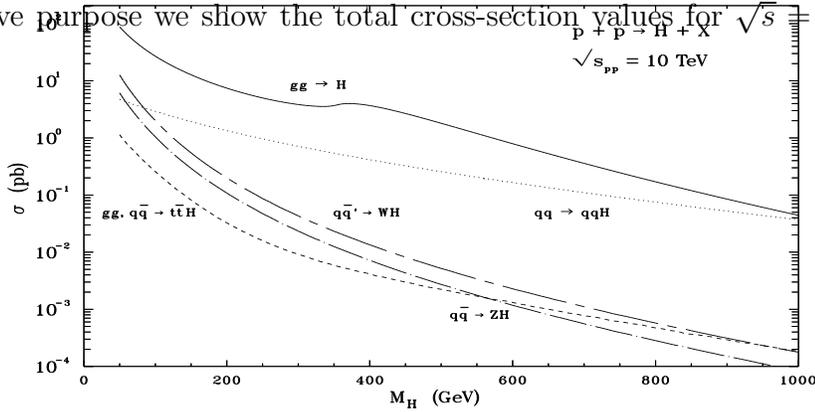}
\end{picture}
\caption{ Total cross sections for $H$ production at the  LHC as a
function of the Higgs mass $M_H$, as given by the four production
mechanisms illustrated in Fig.~2, at $\sqrt s_{pp}=10$~TeV
}
\label{fig:brlo}
\end{figure}
 
There are various uncertainties in the rates of the above 
processes, although none is particularly large. The most significant
are: (i) the lack of precise knowledge  of the gluon
distribution at small $x$, which is important
for the intermediate-mass Higgs, and (ii)  the effect of
unknown  higher-order perturbative QCD corrections.
One can attempt to quantify the former by using 
recent sets of different
parton distributions  which give
excellent fits to a wide range of deep inelastic scattering data~\cite{MWJSZK}.

The next-to-leading order  QCD corrections  are known
for processes (a), (b) and (c) and are included.
 By far the most important of these
are the corrections to the gluon fusion process (a) which have
been calculated in Ref.~\cite{newKfacgg}.
In the limit  where  the Higgs mass is far below the
$2m_t$ threshold, these corrections are calculable analytically
\cite{Kfacgg1,Kfacgg2,Kfacgg3}.
In fact, it turns  out that the analytic result is a good approximation
over the complete $M_H$ range~\cite{DjouadiRev,sdgz}.
In Ref.~\cite{sdgz,MWJSZK} the impact of the next-to-leading order QCD corrections
for the gluon fusion process on LHC cross sections was investigated,
both for the \sm\ and for the \mssm.
Overall, the next-to-leading order correction increases 
the leading-order result by a factor of about 2, when the normalization
and factorization scales are set equal to  $\mu = M_H$.
This `$K$-factor' can be traced to a large constant piece in the
next-to-leading correction \cite{newKfacgg},
\be
\label{Kfac}
K \approx 1 + {\alpha_s(\mu=M_H)\over \pi}\;\left[{\pi^2}+{11\over
2}\right] .
\ee
Such a large  $K$-factor usually implies a non-negligible scale
dependence of the theoretical cross section. 

To judge the quality of the various signals of Higgs production
we must know the event rates for all the promising channels.
Considering all the possible combinations of
production mechanisms and decay channels~\cite{ATLAS,CMS},
the best chance of discovering a \sm\ Higgs at the LHC appears 
to be  given by the 
following signatures: 
(i) $gg\to H\to \gamma\gamma$, (ii)
$q\bar q'\to WH\to \ell\nu_\ell\gamma\gamma$ and (iii) $gg\to H\to
Z^{(*)}Z^{(*)}\to \ell^+\ell^-\ell'^{+}\ell'^{-}$, where $\ell,\ell'=e$
or $\mu$.
Recently, the importance of several other modes has   been
emphasized.  By exploiting techniques of
flavor identification of $b$-jets, thereby reducing the huge QCD
background from light-quark and gluon jets, the modes
 (iv) $q\bar q'\to WH\to \ell\nu_\ell b\bar
b$ and (v) $gg,q\bar q\to t\bar t H\to b\bar b b\bar b WW
\to b\bar b b\bar b  \ell\nu_\ell X$, can be used to search 
for the \sm\ Higgs \cite{Tevadetect1,Tevadetect2}. 
Another potentially important channel, particularly
for the mass range $2 M_W \Ord M_H  \Ord 2 M_Z$,  is
(vi) $ H\to W^{(*)}W^{(*)}\to \ell^+\nu_\ell \ell'^-\bar\nu_{\ell'}$
 \cite{dittdrei}. Here the lack of a measurable narrow resonant
peak is compensated by a relatively large branching ratio,
since for this mass range $H\to WW$ is the dominant decay mode.
Again for sake of illustration we show in Fig.\,4 Higgs production
times branching ratios for various decay modes at two different
energies.
 \begin{figure}[hbtp]
\vspace{7.5cm}
\begin{picture}(7,7)
\includegraphics{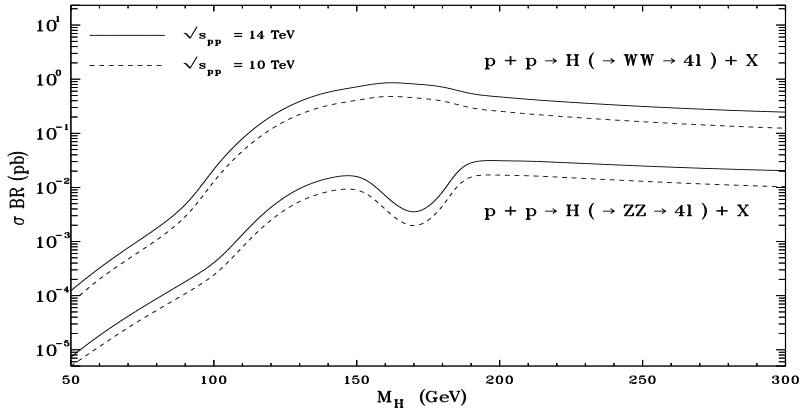}
\end{picture}
\caption{ 
Higgs production times the branching
ratio for the decay modes
 $H\to W^{(*)} W^{(*)} \to 
\ell^+\nu_\ell\ell'^-\bar\nu_{\ell'}$ ($\ell,\ell'=e,\mu$) and
 $H\to Z^{(*)} Z^{(*)} \to 
\ell^+\ell^-\ell'^+\ell'^-\, $
  as a function
of the Higgs mass in the range
  $0 \leq M_H \leq  300$~GeV,
at $\sqrt s_{pp}=10$ TeV and $\sqrt s_{pp}=14$~TeV.
Note that  $m_t=175$ GeV.
}
\label{fig:brlo}
\end{figure}

\section{Conclusion}
The   decay modes
and the production cross sections of the 
most important production mechanisms of the \sm\ Higgs at the
LHC can be calculated with good precision and they are available 
in the literature.
As the most promising
signatures which should allow for Higgs detection at the LHC are
\begin{itemize}
\item $gg\to H\to \gamma\gamma$,
\item $q\bar q'\to WH\to \ell\nu_\ell\gamma\gamma$
and $gg,q\bar q\to t\bar t H\to \ell\nu_\ell\gamma\gamma X$,
\item $q\bar q'\to WH\to \ell\nu_\ell b\bar
b$ and $gg,q\bar q\to t\bar t H\to b\bar b b\bar b WW
\to b\bar b b\bar b  \ell\nu_\ell X$,
\item $gg\to H\to
Z^{(*)}Z^{(*)}\to \ell^+\ell^-\ell'^{+}\ell'^{-}$, where
$\ell,\ell'=e$ or $\mu$,
\item $gg\to H\to
W^{(*)}W^{(*)}\to \ell^+\nu_\ell \ell'^{-}\bar\nu_{\ell'}$,
where $\ell,\ell'=e$ or $\mu$,\footnote{In the analysis of
Ref.~\cite{dittdrei} the additional decay channels $W\to \tau \nu
\to (e,\mu)+\nu$'s were also included, yielding a slightly larger
signal event rate.}
\item $gg\to H\to ZZ\to \ell^+\ell^-\nu_{\ell'}\bar\nu_{\ell'}$,
where $\ell=e$ or $\mu$ and $\ell'=e,\mu$ or $\tau$,
\end{itemize}
The current theoretical errors are estimated  to 
be $\approx \pm 20\%$ (for the uncertainty
due to parton distributions and $\alpha_s$) and 
$\approx\pm 30\%$ (for the error due to the scale dependence, see also
Ref.~\cite{sdgz}),
the latter for the gluon-gluon fusion process.

The theoretical status concerning the physics signal of the branching
ratios and production cross sections
of the Higgs bosons in the MSSM is comparably good.
However, important  QCD corrections
to the cross section values of the  various background processes are
still lacking.
In the case of the supersymmetric Higgs-search, it would be of great significance
to find alternative scenarios to the Higgs sector
of the MSSM.


\section{Acknowledgments}
I thank Stefano Moretti and James Stirling for discussions and
comments, and I also thank
  Bernd Kniehl for organizing a very pleasant workshop.


\section{References}

\begin{thebibliography}{99}

\bibitem{guide} J.F.~Gunion, H.E.~Haber, G.L.~Kane and S.~Dawson, 
                {\it ``The Higgs Hunter Guide''} 
                (Addison-Wesley, Reading MA, 1990).

\bibitem{LHC} Proceedings of the ``{\it Large Hadron Collider Workshop}'', Aachen, 4-9 October
              1990, eds. G.~Jarlskog and D.~Rein, Report CERN 90-10, ECFA 90-133, Geneva, 1990.  

\bibitem{ATLAS} ATLAS Technical Proposal, CERN/LHC/94-43 LHCC/P2 (December 1994).

\bibitem{CMS} CMS Technical Proposal, CERN/LHC/94-43 LHCC/P1 (December 1994).
                       1991 440.

\bibitem{oldpaper} Z.~Kunszt and W.J.~Stirling, in Ref.~\cite{LHC}.

\bibitem{MWJSZK}
 Z. Kunszt, S. Moretti and W.J. Stirling, DFTT-34-95, Nov 1996,
 hep-ph/9611397  
\bibitem{JGSWS}
 J. F. Gunion, A. Stange (BNL) and S. Willenbrock (U. Ill., Urbana),
hep-ph/9602238 (1996)

\bibitem{richter}
D. Froidevaux, 
 Elzbieta Richter-Was,    CERN-TH-96-111 ( 1996).
\bibitem{spirahgludec} M. Spira, CERN-TH/96-285 and references therein

\bibitem{spirasoft}
M. Kr\"amer, E. Laenen and M. Spira, CERN preprint CERN-Th/96-231,
  hep-ph/9611272.
 \bibitem{mssmall}
see \eg 
 B. Kileng, P. Osland and P. N. Pandita
\zp{ C71}{1996}{ 87};
 G. L. Kane, G. D. Kribs, S. P. Martin and J. D. Wells
\pr {D53} {1996}{ 213}, hep-ph/9508265;
 H. E. Haber, R.    Hempfling and A. H. Hoang,
hep-ph/9609331; 
 A. Bartl, H. Eberl, K. Hidaka, T. Kon, W. Majerotto,
   Y. Yamada,
hep-ph/9701398; 
Bartl \etal, Gunion \etal, Dawson, Djouadi, Spira
 S. Dawson, A. Djouadi, M. Spira
\prl {77} {1996}{ 16}


\bibitem{SMHVp} T.~Inami, T.~Kubota and Y.~Okada, Z. Phys. {\bf C18}
                        (1983) 69;
                        H.~Zheng and D.~Wu, \pr D42 1990 3760;
                        A.~Djouadi, M.~Spira, J.J.~van~der~Bij and P.M.~Zerwas,
                        \pl B257 1991 187;
                        A.~Djouadi, M.~Spira and P.M.~Zerwas,
                        \pl B276 1992 350;
                        S.~Dawson and R.P.~Kauffman, \pr D47 1993 1264.

\bibitem{SMHggQCDcorr} 
                       M.~Djouadi, M.~Spira and P.M.~Zerwas, \pl B264

\bibitem{corrreview} B.~Kniehl, DESY Report No.~93-069; 
        M.~Carena, P.M.~Zerwas {\it et al.},
                     in ``Physics at LEP2", eds. G.~Altarelli et.al., 
                     CERN Report 96-01, Vol.1, p.351 (1996). 
\bibitem{threshold} K.~Melnikov, M.~Spira and O.~Yakovlev, 
                    {\it Z. Phys.} {\bf C64} (1994) 401.

\bibitem{MSSMpaper} S.~Moretti and W.J.~Stirling, 
{\it Phys. Lett.} {\bf B347} (1995) 291; Erratum, {\it ibidem} {\bf B366}
(1996) 451.

\bibitem{xggh} H.~Georgi, S.L.~Glashow, M.E.~Macahek and D.V.~Nanopoulos,
               \prl 40 1978 692.
               
\bibitem{xvvh} R.N.~Cahn and S.~Dawson, \pl B136 1984 196.
   
\bibitem{xvh} S.L.~Glashow, D.V.~Nanopoulos and A.~Yildiz, \pr D18
1978 1724;   Z.~Kunszt, Z.~Trocsanyi and W.J.~Stirling, \pl B271 1991
247.

\bibitem{xtth} Z.~Kunszt, \np B247 1984 339; 
               J.F.~Gunion, \pl B253 1991 269; 
               W.J.~Marciano and F.E.~Paige, \prl 66 1991 2433; 
               J.F.~Gunion, H.E.~Haber, F.E.~Paige, W.-K. Tung and
               S.S.D. Willenbrock, \np B294 1987 621; 
               D.A.~Dicus and  S.S.D. Willenbrock, \pr D39 1989 751.        
\bibitem{newKfacgg} D.~Graudenz, M.~Spira and P.M.~Zerwas, \prl 70 1993 1372.

\bibitem{Kfacgg1} A.~Djouadi, M.~Spira and P.M.~Zerwas, 
                  in Ref.~\cite{SMHggQCDcorr}.

\bibitem{Kfacgg2} S.~Dawson, \np B359 1991 283.

\bibitem{Kfacgg3} S.~Dawson and R.P.~Kauffman, \pr D49 1993 2298.

\bibitem{DjouadiRev} A.~Djouadi, {\it Int. J. Mod. Phys.} {\bf A10}
                     (1995) 1; 

\bibitem{sdgz}
M.~Spira, A.~Djouadi, D.~Graudenz and P.M.~Zerwas, \np B453 1995 17.

\bibitem{Tevadetect1} A.~Stange, W.~Marciano and S.~Willenbrock, \pr
D49 1994 1354;~{\it Phys. Rev.}~{\bf D50}~(1994)~4491.

\bibitem{Tevadetect2} J.F.~Gunion and T.~Han, \pr D51 1995 1051.

\bibitem{dittdrei} M.~Dittmar and H.~Dreiner, preprint\ RAL-96-049
(1996), hep-ph/9608317.

\end{thebibliography}
\end{document}